# Effects of Brine Valency and Concentration on Oil Displacement by Spontaneous Imbibition: An Interplay between Wettability Alteration and Reduction in the Oil-Brine Interfacial Tension

Anupong Sukee[1], Tanakon Nunta[1], Nawamin Fongkham[1], Hutthapong Yoosook[2], Montri Jeennakorn[2], David Harbottle[3], Nipada Santha[4], and Suparit Tangparitkul[1,*]

[1]Department of Mining and Petroleum Engineering, Faculty of Engineering, Chiang Mai University, Chiang Mai, 50200, Thailand

[2]Northen Petroleum Development Center, Defence Energy Department, Ministry of Defence, Chiang Mai, 50110, Thailand

[3]School of Chemical and Process Engineering, University of Leeds, Leeds, LS2 9JT, UK

[4]Department of Geological Sciences, Faculty of Science, Chiang Mai University, Chiang Mai, 50200, Thailand

*To whom correspondence should be addressed: Email suparit.t@cmu.ac.th Phone: +66 5394 4128 Ext. 119

## Abstract

Brine fluids have recently been of high interest to enhanced oil recovery process in both academia and industry. Both diluted formation brines and specifically formulated brines were reported to improve crude oil displacement in porous rock, owing to either their bulk salinity or brine-type. Mechanisms for such an improvement were widely proposed, including microscopic interfacial phenomena: wettability alteration and reduction in the oil-brine interfacial tension ($\sigma$), although their synergistic or interlinked contributions were vaguely clarified. To elucidate insights into this "low-salinity" enhanced oil recovery, crude oil displacement by spontaneous imbibition was conducted in the current research with focus on the effects of brine valency and concentration. Monovalent (NaCl) and divalent ($CaCl_2$) brines at elevated concentrations (10, 100, and 1000 mM) were examined as imbibing fluids. Changes in the three-phase contact angle and the crude oil-brine interfacial tension were also investigated. Imbibition results showed that NaCl brine at 'suitable' concentration (100 mM) displaced greater oil (95.8%) than too-low (10 mM) or too-high (1000 mM) concentrations, and these monovalent brines displaced more effective than those of divalent $CaCl_2$ due to an oil-wetting as a result of divalent ion bridging phenomenon. This echoes crucial influences of both brine valency and concentration. Since no direct individual contribution from either the contact angle or $\sigma$ on the oil displacement was obtained, an interplay between these two parameters were thought to control. The imbibition results were a capillary-dominated process (capillary number $< 2.1 \times 10^{-6}$), re-confirmed by their correlations with the calculated capillary pressures and inverse Bond numbers. The findings revealed that in a given imbibition system a required $\sigma$ is a wettability-dependent: water-wet system needs high $\sigma$ to enhance a driving capillarity while oil-wet system prefers lower $\sigma$ to weaken a resisting capillary force. Brine formula directly attributed to wettability and $\sigma$: NaCl brines secure water-wetting with high $\sigma$ while $CaCl_2$ brines reduced $\sigma$ more effectively with an assured oil-wetting. Low-salinity enhanced oil recovery mechanism was thus found to be contributed from capillary effect, which was an interplay between the interfacial tension and wettability. Paring these two parameters by formulating imbibing brine to anticipate high oil recovery is crucial and of challenge.

**Keywords:** spontaneous imbibition; low-salinity waterflooding; capillary; enhanced oil recovery; wettability alteration; interfacial tension

# 1 Introduction

Brine solutions of various formulae have been increasingly implemented in oilfields as displacing fluids for incremental oil displacement process [1]–[3], namely low-salinity enhanced oil recovery (EOR). Since 1990s, either brines diluted from connate water (salinity focused) or engineered brines with selective compositions added or deducted (brine-type concerned) were widely researched owing to their greener and cheaper than other conventional chemical EOR [4]–[7]. Although great efforts have been dedicated to understanding the mechanism and performance of low-salinity EOR from microscopic scale to field trials, dominant mechanisms and their contributions to oil displacement are barely conclusive due to variations in reservoir rock types and brine fluids compared [8]–[10]. To date, a few microscopic low-salinity EOR mechanisms have been suggested, including reduction in the oil-water interfacial tension ($\sigma$), wettability alteration, mineral dissolution, fines migration, emulsification, and electrical double-layer expansion [7], [11]–[15], while their synergistic contribution or governing mechanism on the oil displacement has yet to be explicitly identified.

Previous studies have attempted to establish or justify some of the proposed mechanisms that were attributed from brine fluids on oil displacement performance. With focus on entire brine concentration (no brine-type considered), there is an 'optimal' salinity that yields the highest oil displacement for a given rock-oil-brine system. Diluted seawater composed of various brine species was usually examined its potential to oil recovery, and research found that two- to four-times dilutions (~8000 – 20000 ppm) were roughly optimal to produce the highest oil recovery [4], [5]. Based on published studies, a few EOR mechanisms were concurrently claimed to improve the oil displacement, consisting of wettability alteration toward water-wet, reduction in $\sigma$, and even dissolution of crude oil polar components in brine [4], [5].

With systematical focus on brine type or valency effect, monovalent brines (i.e. NaCl) was found to displace greater oil than multivalent brines (e.g. divalent $CaCl_2$) at equivalent concentration. Such performance was mainly associated with changes in $\sigma$ and wettability [7], [15], [16]. However, when considers these microscopic mechanisms contributed from two types of brine, their contributions are inconsistent or even conflict and hence designing brine fluids for EOR are such a dilemmatic challenge [17]–[19]. Recent works confirms that monovalent brines could promote a strong water-wet system due to repulsive hydration forces constructed between oil-brine and brine-rock interfaces [15], [20] rather than multivalent brines that tend to bridge crude oil components onto rock surfaces [21], [22], while divalent brines were found to be more effective than monovalent species on reducing the $\sigma$ due to a stronger screening of electrostatic repulsion between polar species at the oil-water interface [23]

Research on NaCl and CaCl$_2$ brines highlighted their distinguish contributions to spontaneous imbibition process, which low-concentration NaCl brines displaced highest oil recoveries [16], [24]. The studies concluded that changes in wettability and $\sigma$ were EOR mechanisms, and a role of resulted capillary force was discussed to control the displacement behavior [25]. However, some studies suggested that the electrical double-layer expansion or increased repulsive electrostatic forces between oil-brine and brine-rock interfaces was attributed to improved oil displacement in waterflooding experiment, though monovalent NaCl brine was endorsed as highly effective displacing fluid rather than other divalent brines [6], [7]. Previous research that focused on effects of brine valency and concentration on oil displacement are summarized in Table 1.

According to our previous work [15], we have elucidated the underlying microscopic effects of low-salinity brines, highlighting a dramatic change in wettability that depended on both brine valency and concentration, while a negligible effect on reduction in $\sigma$ was observed. The current research is a continuation from such a previous once by means of a systematic set of spontaneous imbibition experiments, with aim to further address the direct influences of brine valency and concentration in 'core-scale' displacement [8]. Brine fluids used in the current work are monovalent NaCl and divalent CaCl$_2$ at different concentrations that representing ranges of low-salinity brine toward formation water (10, 100, and 1000 mM). Two microscopic properties, namely the three-phase contact angle ($\theta$) and the crude oil-brine interfacial tension, were measured. Dimensionless numbers and capillary pressure were also calculated to describe the insights into imbibition behavior and oil displacement results.

**Table 1.** Previous research on effects of brine valency and concentration on oil displacement.

| Displacement scheme | Brine fluids | Porous rock and oil phase | Remarked results and EOR mechanisms | Reference |
|---|---|---|---|---|
| Spontaneous imbibition | NaCl and CaCl$_2$ (1 – 15 wt%) | Carbonate and crude oil | - 2.5 wt% NaCl and 5 wt% CaCl$_2$ brines produced incremental oil of 8.1% and 7.3%, respectively, implying dependences on brine valency and concentration<br>- EOR mechanisms were dominantly wettability alteration (toward water-wet) and additional reduction in $\sigma$ | Valluri et al. [16] |
| Spontaneous imbibition | NaCl, KCl, MgCl$_2$, Na$_2$SO$_4$, and CaCl$_2$ (500 – 100000 mg/L) | Sandstone and crude oil mixed with kerosene | - Monovalent NaCl brine produced the highest oil recovery of 35.4% at 5000 mg/L<br>- Greater imbibition results were contributed from enhanced hydrophilicity of the rock<br>- EOR mechanism was a combination of low $\sigma$ and substantial wettability alteration toward hydrophilicity | Zhu et al. [24] |
| Waterflooding | NaCl, CaCl$_2$, and MgCl$_2$ (2000 – 50000 mg/L) | Sandstone and crude oil | - Monovalent NaCl brine produced the highest oil recovery of 84.8% at 2000 mg/L<br>- EOR mechanism was attributed to electrical double layer expansion with higher repulsive force between interfaces, thus wettability alteration (toward water-wet) | Nasralla et al. [6] |
| Waterflooding | KCl and CaCl$_2$ (0.1 – 1 wt%) | Model rock (mixture of carbonate and sandstone) and crude oil | - Monovalent KCl brine produced the highest oil recovery of 61% (0.25 wt%), while divalent CaCl$_2$ brine produced slightly lower oil recovery of 59% (0.1 wt%)<br>- EOR mechanisms were a reduction in $\sigma$ and increase in repulsive electrostatic force between mineral-brine and oil-brine interfaces | Debnath et al. [7] |

## 2 Materials and Experimental Methods

### 2.1 Crude oil and rock samples

Crude oil collected from a primary-production well (FA-BT63-10) at Fang oilfields in Chiang Mai (Thailand) was used throughout the study as an oil phase. The crude oil was shaken and de-gassed before use. The dead crude oil has density of 854.2 kg/m$^3$ and viscosity of 18 mPa·s at 50 °C. The crude oil has high wax content of 51.0 wt% with slight asphaltenes of 0.05 wt%.

Rock core samples (8.5 cm diameter and 15 cm length cylinders, Fig. 1a) were drilled from a single block of sandstone outcrop, taken from Doi Noi in Chiang Mai (Thailand), to ensure that the rock properties are consistent for the imbibition study. Thin slice samples of the rock (~3 mm thickness, Fig. 1a) were also prepared for the measurement of the three-phase contact angle (to be discussed in the section below). The rock samples were preliminarily cleaned by saturating with deionized water and left to dry overnight in an oven. Microscopic image of the rock thin section (Fig. 1b) was used to determine pore size using an image processing program (ImageJ). The average pore size was analyzed to be 240 ± 5 μm. It is noted that only one rock species of sandstone was considered in the current study, and mechanisms or phenomena mentioned in the following discussions are mainly associated with sandstone.

Effective porosity of the core samples was determined by liquid saturation method using deionized water. The porosity of all core samples was consistent of ~4.0%, a fairly low porosity for clastic rock, which capillary force normally controls fluid flow behavior [26].

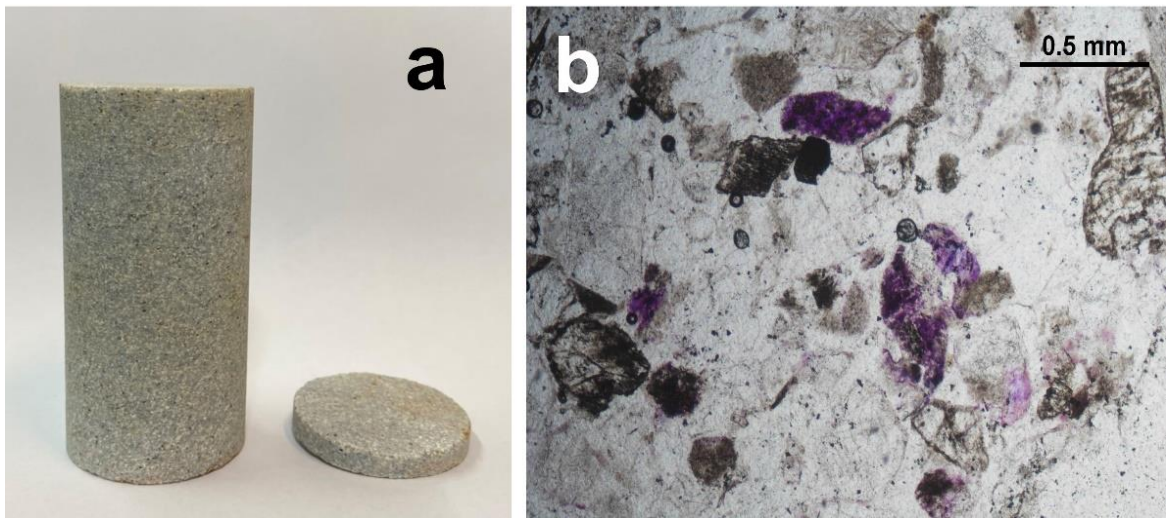

**Figure 1.** Sandstone core and thin slice samples (a) and thin-section image for pore (color-dyed) size analysis (b).

## 2.2 Brine fluids

Two brine types (i.e. monovalent and divalent salts) were compared concurrently with an effect of brine concentration in three orders of magnitude that are relevant to petroleum formation and displacing fluids (i.e. 10, 100, 1000 mM). Sodium chloride as monovalent salt (NaCl ≥ 99.5%, RCI Labscan, Thailand) and calcium chloride dehydrate as divalent salt ($CaCl_2$ ≥ 99.5%, RCI Labscan, Thailand) were used as received to prepare the brine solutions using deionized water (RCI Labscan, Thailand), see Table 2. Synthetic formation brine that has equivalent salt compositions (composed of both salt types) to actual formation water collected from Fang oilfields was also considered in the current study.

**Table 2.** Brine compositions used throughout the study.

| Brine fluids | Cations (ppm) | | Anion (ppm) | | | Total Dissolved Solids (ppm) |
|---|---|---|---|---|---|---|
| | $Na^+$ | $Ca^{2+}$ | $Cl^-$ from $Na^+$ | $Cl^-$ from $Ca^{2+}$ | Total $Cl^-$ | |
| Formation brine (13.42 mM) | 292 | 30 | 449 | 53 | 502 | 824 |
| 10 mM NaCl | 292 | - | 292 | - | 292 | 584 |
| 100 mM NaCl | 2,920 | - | 2,920 | - | 2,920 | 5,840 |
| 1000 mM NaCl | 29,200 | - | 29,200 | - | 29,200 | 58,400 |
| 10 mM $CaCl_2$ | - | 370 | - | 740 | 740 | 1,110 |
| 100 mM $CaCl_2$ | - | 3,700 | - | 7,400 | 7,400 | 11,100 |
| 1000 mM $CaCl_2$ | - | 37,000 | - | 74,000 | 74,000 | 111,000 |

## 2.3 Measurements of the crude oil-water interfacial tension and the three-phase contact angle

The interfacial tension between the crude oil and brine solution ($\sigma$) was measured by a pendent drop method using an Attension® Theta Optical Tensiometer (TF300-Basic, Biolin Scientific, Finland) equipped with a thermal cell (C217W, Biolin Scientific, Finland) and operated at 50 ± 1 °C. A 10 μL droplet of crude oil was dispensed at the tip of a stainless inverted needle (gauge 22) using a micro-syringe pump (C201, Biolin Scientific, Finland). The shape of the oil droplet was monitored at 3.3 fps until no detectable change was observed (~3600 s).

An Attension® Theta Optical Tensiometer (TF300-Basic, Biolin Scientific, Finland) was also used to measure the three-phase oil-brine-rock contact angle ($\theta$) at 50 ± 1 °C by an inverted sessile drop method on aged rock thin slice. The cleaned thin slice was aged with the formation water and crude oil as same as the core samples used for the imbibition experiment described in the next section. The aged thin slice was stationed on a home-made acrylic stand placed inside a thermal cell (C217W, Biolin Scientific, Finland), with the rock thin slice hang in a center of the cell window for contact angle visualization. Brine solution was pre-heated to 50 °C and added to the thermal cell to submerge the thin slice. An oil droplet of ~5 µL was deposited underneath the thin slice by means of an inverted needle attached to a micro-syringe, and the oil-brine-rock contact angle was formed. The contact angle was recorded at 3.3 fps until the steady-state contact angle (no detectable change) was attained (~1 h). The three-phase contact angle was measured through the brine phase from image analysis using the OneAttention software. The average contact angle from the left and right is reported.

For both measurements, the brine solutions and formation brine in Table 2 were examined. The measurements were conducted in triplicate and the average values were reported and used for further analyses.

**2.4 Spontaneous imbibition experiment**

Prior to imbibition experiment, the cleaned core samples were saturated with formation water for 2 days at 50 °C and left to dry in an oven for 2 days to reconstruct connate salts deposited in pristine cores. The core samples were thereafter saturated with crude oil and aged at 50 °C for 7 days. An initial oil saturation in each core sample was determined by means of weight difference after the aging process. Due to direct adsorption of crude oil components and co-adsorption via salt ions on rock surface, the aging process has established an initial wettability of core samples to be weakly water-wet as characterized by the measured three-phase contact angle of 66.0°.

For spontaneous imbibition experiment, a glass Amott cell was cleaned before each use by rinsing thoroughly with toluene and left dry overnight. An oil-saturated core sample was placed in the Amott cell chamber before the cell was steadily filled with imbibing fluid (brine fluid) up to a scale reading of graduated cylinder (Fig. 2). The core sample was thus fully submerged in an imbibing fluid. To induce a co-current flow, a 1-cm-height stainless-wire truss was used to partition a core sample from the Amott cell bottom to allow a vertical flow from bottom-end of the core sample [27].

All imbibition experiments were conducted in an oven at operating temperature of 50 °C, see Fig. 2a. The Amott cell and imbibing fluids were pre-heated in the oven at 50 °C before use. Fluid evaporation from the Amott cell was prevented by sealing the cell and tubing ducts with Parafilm®. Oil produced from a core sample gradually liberated and floated up to accumulate at graduated cylinder scale bar, where the produced oil volume was recorded periodically until no noticeable change in oil volume was observed. Oil recovery is reported in percentage of produced oil volume to initial oil saturation as a function of time. The imbibition experiments were conducted with 6 brine fluids listed in Table 2.

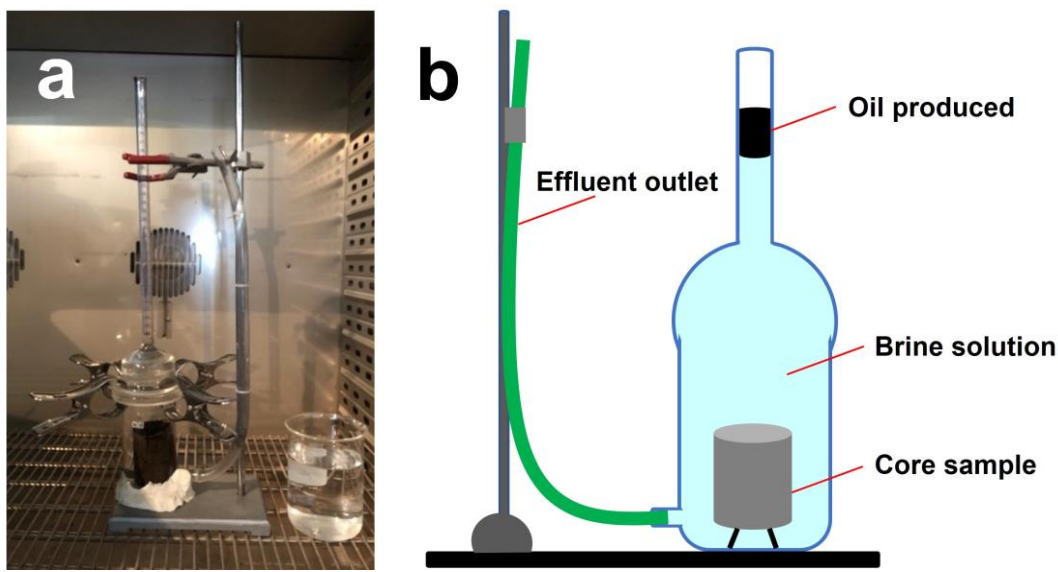

**Figure 2.** Amott cell used in spontaneous imbibition experiment (a) and its schematic diagram showing main components with core sample submerged in brine solution (b).

## 3 Results and Discussion

### 3.1 Crude oil-brine interfacial tension and the three-phase contact angle

The crude oil-brine interfacial tension was measured in NaCl and $CaCl_2$ brine solutions at concentrations of 10, 100, and 1000 mM at 50 °C (Fig. 3a). An increase in the brine concentration resulted in a reduction in the $\sigma$. At the high brine concentration of 1000 mM NaCl, the $\sigma$ was decreased to be 28.7 mN/m when compared to $\sigma$ (35.9 mN/m) at the lower concentration of 10 mM NaCl. Similar trend was even more distinctive with $CaCl_2$ brines. With increasing brine concentration, the $\sigma$ reduction was attributed to an increased adsorption of surface-active species from crude oil components (e.g. naphthenic acids and asphaltenes) [28], [29]. Although the crude oil surface-active species partition at the crude oil-brine interface spontaneously, increasing cations in brine neutralize the charged head groups of anionic oil components and thus induce them to pack much densely at the oil-brine interface [15]. At equivalent concentration, reduction in $\sigma$ was more effective in divalent brines than that of monovalent as observed in our previous study [15]. This is attributed to a stronger screening of electrostatic repulsion between polar species at the oil-brine interface by $Ca^{2+}$ than $Na^+$, which is due to the Ca-ion being strongly hydrated and of higher valency [15], [23], [30].

Even though such reductions in $\sigma$ were substantially distinctive, change in $\sigma$ is usually more effective for EOR as its main mechanism if such a change is in an order of magnitude. This will be justified thoroughly in the later section concurrently with the effect of other mechanism, i.e. wettability alteration.

It is noted that a minimum $\sigma$ and the $\sigma$ reversal behavior (due to reduced adsorption from the salting-out effect [28], [31]) have not been observed for both brine species in the current study as observed in our previous study [15]. It is thought that such a behavior would be attained at much higher brine concentration since the crude oil in the current study likely composes of more effective surface-active species than the heavy crude oil used in the previous work [15]. This was also evidenced by a greater reduction in $\sigma$ with the current oil ($\leq$ 25 mN/m $\Delta\sigma$) than the oil used in the previous study ($\leq$ 2 mN/m $\Delta\sigma$) [15].

Figure 3b shows the measured three-phase contact angle of crude oil droplet at steady state in various brines, which change in the $\theta$ was observed in both monovalent (NaCl) and divalent ($CaCl_2$) brines. With increasing NaCl brine concentration, the $\theta$ decreased from 66.0° of an initial wettability to be 56.0°, 32.0°, and 44.0° for 10, 100, and 1000 mM, respectively. These measured $\theta$ in NaCl brines were less than 90°, indicating a water-wet characteristics. On the contrary, increase in $CaCl_2$ brine concentration resulted in a dramatic increase in $\theta$ to be 76.1°, 126.9°, and 131.1° for 10, 100, and 1000 mM, respectively, reflecting an oil-wet characteristics. The measured $\theta$ in the brine systems are not dominantly controlled by the change in $\sigma$, even though the $\theta$ should be correlated with $\sigma$ as theoretically described by the Young's equation [32]. The current results agree with our previous report [15], which proved that the $\theta$ measured in the brine system is influenced dominantly by the colloidal forces and interactions between the oil-brine and brine-rock interfaces, i.e. the change in the substrate-brine interfacial tension.

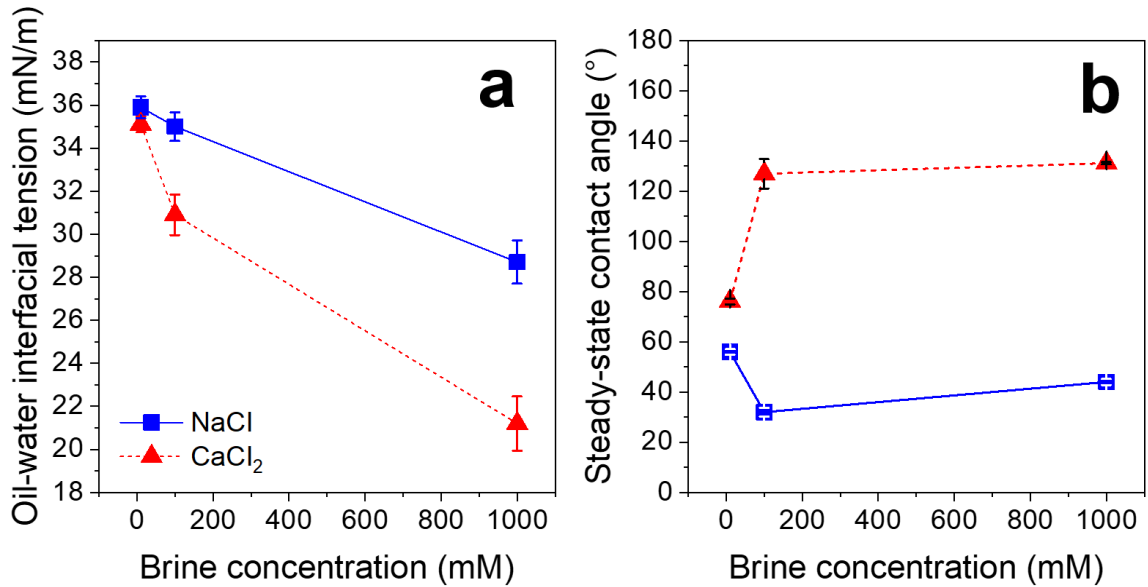

**Figure 3.** Crude oil-brine interfacial tension (a) and the steady-state contact angle of crude oil droplet on rock thin slide in brine solutions (b) as a function of brine type (i.e. NaCl and $CaCl_2$) and brine concentration at 50 °C. Error bars are standard deviation. Lines to guide the eye.

In monovalent NaCl brines, the $\theta$ decrease (i.e. 56.0° to 32.0° with NaCl concentration changed from 10 to 100 mM) was controlled by the repulsive disjoining pressure attributed to a contribution from non-DLVO hydration forces that only exist in brine system [15], [20]. These hydration forces result from hydrated cation that irreversibly adsorbed onto substrate, see Fig. 4a. As a strong and short-range forces, the hydration forces were reported to screen other long-range DLVO (van der Waals and electrostatic) forces [33]. Our results observed a unique behavior attributed to the hydration forces. When increasing NaCl concentration to about 1000 mM order of magnitude, the hydration forces weakened as a result of hydrated cation radius reduction [20] and the total disjoining pressures rather aligned with the theoretical DLVO [30], [34], [35] (i.e. $\theta$ increased from 32.0° to 44.0° with NaCl concentration changed from 100 to 1000 mM).

Change in the measured $\theta$ in divalent $CaCl_2$ brines differed from those of NaCl. Owing to divalent ion bridging of surface-active species (i.e. anionic naphthenic acids) in crude oil adsorbed on rock substrate via divalent $Ca^{2+}$ cation, the interaction is dominated by a long-range attractive hydrophobic forces between the crude oil-brine and substrate-brine interfaces [34], [36]–[38], see Fig. 4b. As such, the hydrophobic attraction is strengthened with increasing $CaCl_2$ concentration and more binding of crude oil surface-active species, and this attributed to an increase in the measured $\theta$ in CaCl$_2$ brines (Fig. 3b). At high concentration ($\geq$ 100 mM CaCl$_2$), the wettability even reversed to be oil-wet with $\theta \gg 90°$.

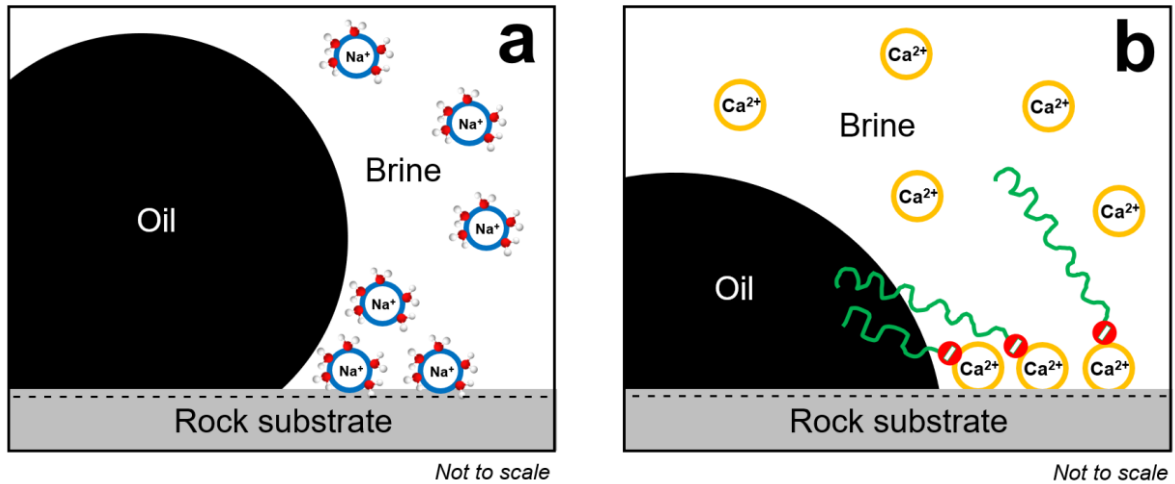

**Figure 4.** Schematic interactions at the crude oil-brine-rock interface that influence the wettability alteration. For monovalent brines, repulsive hydration forces constructed in brine as a result of hydrated cation dominate crude oil droplet displacement from substrate (a). For monovalent brines, adhesive hydrophobic forces between the crude oil-brine and substrate-brine interfaces induced by divalent ion bridging of surface-active species (shown as green chain with anionic head) hinders crude oil displacement from substrate (b).

## 3.2 Spontaneous imbibition result

Spontaneous imbibition from oil-saturated core displaced by different brines were conducted, and dynamic and ultimate oil recovery results are shown in Fig. 5. Generally, crude oil was displaced at large amount at initial stage of imbibition process (at $\leq$ 75 h) and then approached the steady state at the later stage, where insignificant oil was recovered and ultimate oil recovery was defined (~150 h). NaCl brines (Fig. 5a) imbibed into core sample to displace saturated crude oil at faster rates and yielded higher oil recoveries than those of CaCl$_2$ brines (Fig. 5b) at all equivalent brine concentrations (Fig. 5c).

Although it seems that crude oil imbibition was enhanced with increasing NaCl brine concentration (from 10 to 100 mM), the highest NaCl brine concentration (1000 mM) did not displace crude oil at the fastest rate nor the highest ultimate oil recovery. The best NaCl brine concentration that produced the maximum oil recovery was 100 mM, which recovered 95.8% oil at steady state (Fig. 5c), while 10 and 1000 mM brines produced 43.4% and 76.7%, respectively. This suggests that oil recovery from imbibition process in the brine system is not 'monotonically' concentration-dependent.

In $CaCl_2$ brines, crude oil imbibition results were fairly poor and also did not depend on brine concentration. The maximum oil recovery in $CaCl_2$ brine (25.8%) was achieved with 10 mM, the lowest concentration. For the higher $CaCl_2$ brine concentrations, the oil recoveries were likely unnoticeable (0.0% for 100 mM and 6.0% for 1000 mM). Although it is noted that 6.0% oil recovery from the highest concentration of 1000 mM brine might be considered substantial, mechanistic understanding on such a non-monotonic relationship between brine concentration and ultimate oil recovery are inconclusive and to be elucidated below.

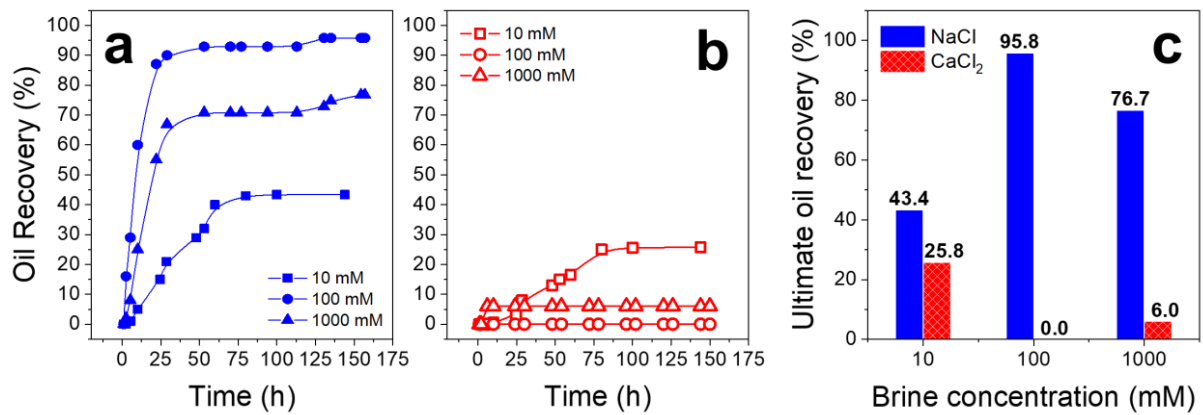

**Figure 5.** Oil recovery from imbibition experiment as a function of time and brine concentration (10, 100, 1000 mM) of NaCl (a) and $CaCl_2$ brines (b). Ultimate oil recovery from both brine types summarized in (c).

Immiscible fluid-fluid displacement occurring in such a spontaneous imbibition process is well established to be controlled by capillarity rather than advection or gravitational effects [27], [39]. With two measurable parameters (i.e. $\sigma$ and $\theta$; Fig. 3) expressing such a dominant capillary effect, the imbibition results (hereafter quantified by residual oil saturation) were therefore plotted against these two parameters to individually validate their contributions, see Fig. 6.

Although the $\sigma$ is a factor that attributes to how strong the capillary force is [27], changes in $\sigma$ do not establish a correlation with residual oil saturation (i.e. the oil displacement) at all in the current brine systems (Fig. 6a). Such an inconsistent was more obvious at $\sigma$ ~35 mN/m, pairing against a wide range of residual oil saturation (~0 – 80%).

On the contrary, a relationship between $\theta$ and residual oil saturation was observed in both oil-wet and water-wet regimes (Fig. 6b). With decreasing $\theta$, the residual oil saturation decreased accordingly and hence a greater oil recovery. This suggests that the $\theta$ is rather a controlling factor via capillary effect for spontaneous imbibition in the brine system.

Effect of brine valency on oil displacement was notably obvious with the measured $\theta$ that grouping into two sets, see Fig. 6b. Monovalent NaCl brines (blue symbols), that secured water-wet $\theta$, contributed to lower residual oil saturation (i.e. greater oil recovery). Divalent $CaCl_2$ brines (red symbols) with resulted $\theta$ toward oil-wet inefficiently displaced crude oil from saturated cores. With effects of brine valency and concentration on wettability (i.e. $\theta$) as discussed in the previous section, Fig. 6b also emphasized their contributions to crude oil displacement.

However, a contribution from $\sigma$ could not be completely neglected since reduced $\sigma$ seems to be desirable for oil displacement in 1000 mM $CaCl_2$ brine. When carefully considered 100 and 1000 mM $CaCl_2$ results, 6.0% oil recovery produced from 1000 mM $CaCl_2$ brine was thought to be attributed from a lower $\sigma$ of 22.3 mN/m (when compared to 100 mM of 35.1 mN/m with 0.0% recovery) rather than their similar $\theta$ (126.9° and 131.1° for 100 and 1000 mM, respectively). This therefore suggests an interplay between these two parameters ($\sigma$ and $\theta$), which will be further justified in the next section.

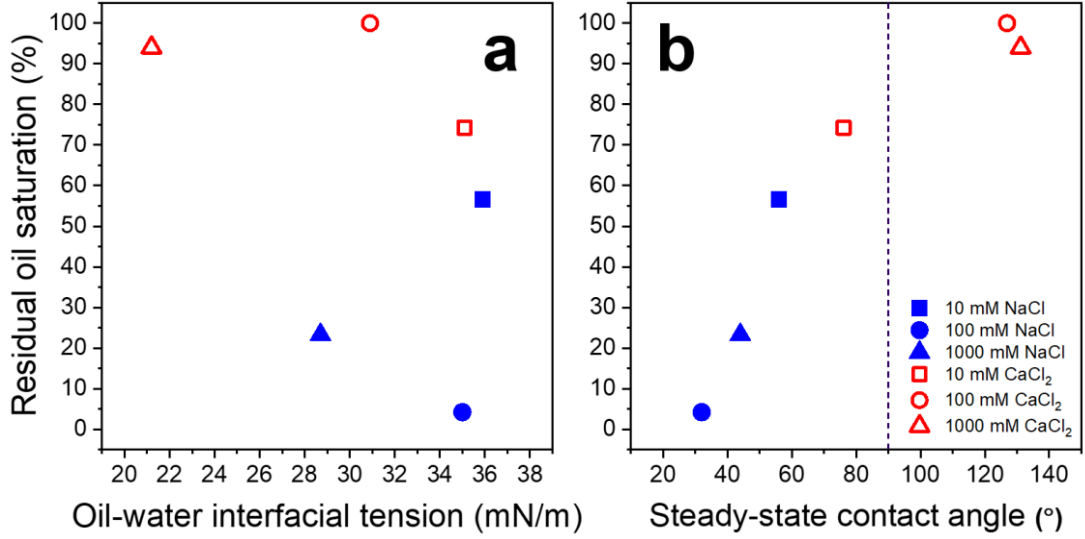

**Figure 6.** Residual oil saturation in core samples after imbibition experiment with both NaCl and CaCl$_2$ brines as a function of the steady-state oil-brine interfacial tension (a) and steady-state contact angle (b). Dash line distinguishes water-wet ($\theta < 90°$) and oil-wet ($\theta > 90°$) regimes.

**3.3 Dimensionless number, capillary pressure, and oil displacement mechanisms**

Since either reduction in the crude oil-brine interfacial tension or change in the contact angle (wettability alteration) could not individually describe the observed imbibition experiments (Fig. 6), a concurrent contribution combined from both parameters is anticipated. Dimensionless numbers, consist of capillary number ($N_{Ca}$) and inverse Bond number ($N_B^{-1}$) [40], [41], that account for both parameters are analyzed. Capillary pressure ($P_c$) with contributions from both $\sigma$ and $\theta$ was also determined to elucidate an interplay between $\sigma$ and $\theta$ on the current oil displacement study by spontaneous imbibition.

Capillary number ($N_{Ca}$) defines as a ratio of viscous to capillary forces:

$$N_{Ca} = \frac{\mu V}{\sigma cos\theta} \quad (1)$$

where $\mu$ is the oil viscosity, and $V$ the characteristic velocity of the crude oil phase.

Inverse Bond number ($N_B^{-1}$) characterizes a ratio between capillary and gravitational forces:

$$N_B^{-1} = \frac{\sigma cos\theta}{r\Delta\rho g H} \quad (2)$$

where $r$ is the average pore radius, $\Delta\rho$ the density difference between crude oil and water phases, $g$ the gravity acceleration constant, and $H$ the core length characteristic taken to be the core height.

Capillary pressure ($P_c$) refers to capillary sucking within pore configuration occurring between two immiscible fluids.

$$P_c = \frac{2\sigma cos\theta}{r} \quad (3)$$

The dimensionless numbers ($N_{Ca}$ and $N_B^{-1}$) and $P_c$ for each imbibition experiment were calculated and reported in Table 3. In all brine fluids, with assuming $V$ of $1 \times 10^{-6}$ m/s the calculated $N_{Ca}$ were substantially low ($< 2.1 \times 10^{-6}$), implying a great capillary control rather than viscous advection [27][42]. Other two quantities ($N_B^{-1}$ and $P_c$) have a dramatic variation for different brine fluids, especially being negative in highly concentrated divalent brines (i.e. 100 and 1000 mM $CaCl_2$). Hence, the calculated $N_B^{-1}$ and $P_c$ were plotted against residual oil saturation to examine their contribution to the oil displacement, shown in Fig. 7.

**Table 3.** Calculated capillary number ($N_{Ca}$), inverse Bond number ($N_B^{-1}$), and capillary pressure ($P_c$) for each brine fluid in spontaneous imbibition experiments.

| Brine fluids | $\sigma$ (mN/m) | $\theta$ (°) | $N_{Ca}$ | $N_B^{-1}$ | $P_c$ (Pa) |
|---|---|---|---|---|---|
| 10 mM NaCl | 35.9 | 56.0 | $9.0 \times 10^{-7}$ | 1.55 | 334.9 |
| 100 mM NaCl | 35.0 | 32.0 | $6.1 \times 10^{-7}$ | 2.27 | 494.8 |
| 1000 mM NaCl | 28.7 | 44.0 | $8.7 \times 10^{-7}$ | 1.39 | 343.6 |
| 10 mM $CaCl_2$ | 35.1 | 76.1 | $2.1 \times 10^{-6}$ | 0.65 | 140.2 |
| 100 mM $CaCl_2$ | 30.9 | 126.9 | $9.7 \times 10^{-7}$ | -1.41 | -309.4 |
| 1000 mM $CaCl_2$ | 21.2 | 131.1 | $1.3 \times 10^{-6}$ | -1.06 | -232.1 |

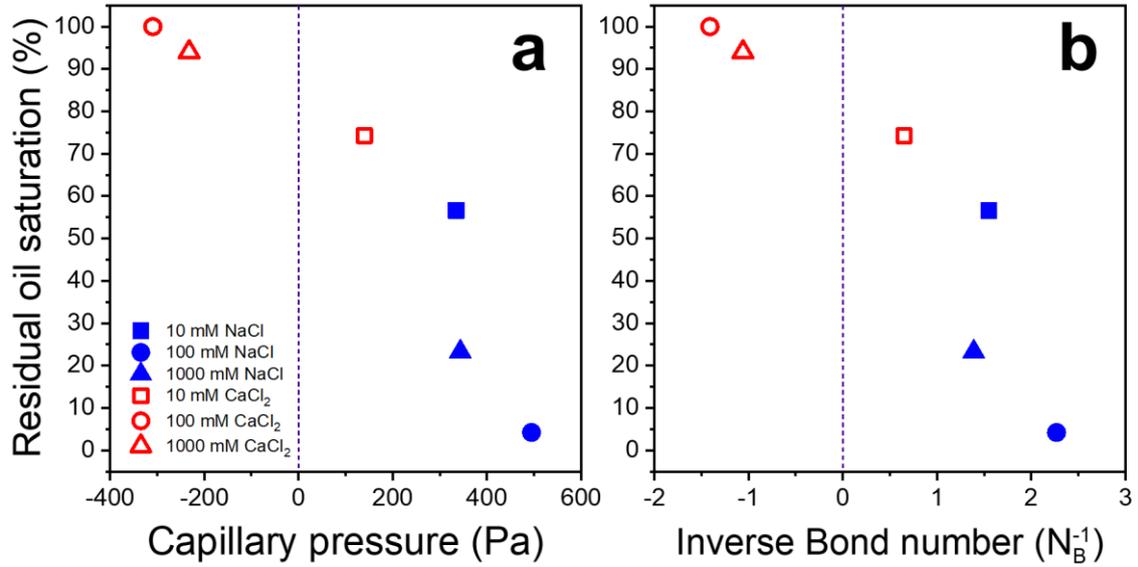

**Figure 7.** Residual oil saturation in core samples after imbibition experiment with both NaCl and CaCl$_2$ brines as a function of $P_c$ (a) and $N_B^{-1}$ (b). Dash lines separate negative and positive regimes.

Good correlations with residual oil saturation were observed for both $P_c$ and $N_B^{-1}$ in Fig. 7, suggesting their strong influences on the oil displacement by the current imbibition study. Residual oil saturation decreased with increasing in both $P_c$ and $N_B^{-1}$. Since capillary force controlled the oil displacement as discussed previously ($N_{Ca} < 2.1 \times 10^{-6}$), the capillary pressure ($P_c$) quantified is thus anticipated to amplify such effect. Calculated $P_c$ values were in both positive and negative regimes (Fig. 7a), which physically interpret as driving and resisting capillary forces, respectively [27][41]. Such positivity and negativity of the values were directly attributed from a cosine function of the contact angle measured ($cos\theta$), or effectively a 'wettability control' [43].

In NaCl brines, the $P_c$ were all positive and relatively high, and hence a great driving capillarity displaced crude oil from porous rock and resulted in high oil recovery (low residual oil saturation). Especially in 100 mM NaCl brine, the highest $P_c$ (494.8 Pa) attributed to the highest oil recovery (95.8%), see Fig. 5c. It is noted that increase in $\sigma$ also contributed to strengthen a driving capillary force in these monovalent brines, and thus benefited the oil recovery.

For CaCl$_2$ brines, the $P_c$ were relatively low or even reversed to be negative (at ≥ 100 mM) owing to wettability alteration as discussed in previous section, see Fig. 3b. At high salinity (100 and 1000 mM), the capillary forces were resistance and hindered the oil displacement (negative $P_c$ due to $\theta > 90°$), which resulted in high residual oil saturation. Interestingly, the $P_c$ value in 1000 mM CaCl$_2$ brine was higher than that of 100 mM CaCl$_2$ (Table 2), and hence slight oil recovered from 1000 mM CaCl$_2$ brine (6.0%) despite their equivalent $\theta$. This was attributed to a substantially reduced or lower $\sigma$ (30.9 mN/m → 21.2 mN/m) that improves the oil displacement in oil-wet regime (i.e. a resisting capillarity), which opposes to the water-wet regime that constructs a driving capillary as discussed above.

Regarding $N_B^{-1}$ correlation shown in Fig. 7b, general observation is similar to that of $P_c$. The positivity and negativity of $N_B^{-1}$ (and $P_c$) were due to $cos\theta$ value, which requires different favorable $\sigma$ value to enhance oil displacement. Absolute values of $N_B^{-1}$ were greater than 1 in most of the brine fluids examined, indicating a relatively weak gravitational contribution to such an imbibition process as also reported in previous works [27][41], [44]. Excepting 10 mM CaCl$_2$ brine, the absolute $N_B^{-1}$ was slightly less than 1 due to less $cos\theta$ value, suggesting some weak gravitational contribution at play.

With $P_c$ and $N_B^{-1}$ analyses, the oil displacement in a spontaneous imbibition process was mainly controlled by capillarity via an interplay between the two interfacial parameters: $\theta$ and $\sigma$. Modifications to these parameters are often highlighted as EOR mechanisms, namely wettability alteration and reduction in the oil-water interfacial tension, respectively [27], [41], [45], [46]. As revealed from the current study, required $\sigma$ value depends on wettability ($\theta$) regimes: water-wet system needs high $\sigma$ while oil-wet system prefers lower $\sigma$ to effectively displace oil. Hence, the $\sigma$ contribution to oil recovery could not be ignored over the wettability prime.

Brine valency and concentration directly influence wettability of crude oil-brine-rock system ($\theta$) and thus indicates which sort of $\sigma$ preferred, although the $\sigma$ is also changed inevitably upon changes in brine valency and concentration (Fig. 3). Pairing of these two parameters in brine fluids (i.e. smart water or low-salinity brine) to yield great oil recovery is crucial and has been illustrated in the current experimental study.

## 4 Conclusion

The current study has demonstrated how the interfacial phenomena (i.e. the three-phase contact angle and the crude oil-brine interfacial tension) have an interplay or influence on the crude oil displacement by spontaneous imbibition process. Effects of brine valency and concentration on the oil displacement were highlighted, with monovalent NaCl and divalent CaCl$_2$ at elevated concentrations. Based on the experimental results with dimensionless numbers and capillary pressure analyzed, the mechanisms for low-salinity EOR linking micro-scale basis to core-scale observation are as follows:

(i) Increasing brine concentration generally induced further reduction in the $\sigma$. Significant reduction effect was observed in $CaCl_2$ brines due to a stronger screening of electrostatic repulsion at the crude oil-brine interface than NaCl brines. Their effects on wettability alteration differed. With increasing brine concentration, NaCl brines likely secured $\theta$ toward water-wet due to hydration forces constructed while $CaCl_2$ brines resulted in high $\theta$ values (i.e. oil-wet) due to divalent cation bridging.

(ii) According to imbibition results, highest oil recovery was obtained with a suitable NaCl brine concentration (100 mM) that induced the strongest water-wetness (i.e. lowest $\theta$) and relatively high $\sigma$. At equivalent concentrations, NaCl brines displaced crude oil more effective than $CaCl_2$ brines, where crude oil was barely recovered. No direct influence of the measured $\sigma$ or $\theta$ was found to correlate with such imbibition results individually.

(iii) With low capillary number obtained, the current imbibition was controlled by capillary force, and this suggests considering the $\sigma$ and $\theta$ effects concurrently. The capillary pressure and inverse Bond number that include those two parameters were therefore quantified and a good correlation with the imbibition results was found. This suggests that low-salinity EOR mechanism in such a spontaneous imbibition process was robustly controlled by an interplay between changes in the two interfacial phenomena: wettability alteration and reduction in $\sigma$.

**CRediT authorship contribution statement**

**Anupong Sukee:** Visualization, Formal analysis, Writing - Original Draft; **Tanakon Nunta:** Methodology, Investigation, Resources; **Nawamin Fongkham:** Validation, Formal analysis, Investigation; **Hutthapong Yoosook:** Resources; **Montri Jeennakorn:** Resources; **David Harbottle:** Supervision, Funding acquisition; **Nipada Santha:** Conceptualization, Supervision; **Suparit Tangparitkul:** Conceptualization, Methodology, Visualization, Formal analysis, Writing - Review & Editing, Supervision, Project administration, Funding acquisition.

**Acknowledgement**

Financial support for this work is greatly acknowledged with contributions from Office of the Permanent Secretary for Ministry of Higher Education, Science, Research and Innovation (Grant No. RGNS 64 – 081) (S.T.), Thailand Science Research and Innovation (TSRI) (S.T.), and Chiang Mai University. N.F. acknowledges his TA/RA scholarship from the Graduate School, Chiang Mai University and currently holds a Tangparitkul Tricenary Junior Fellowship at Chiang Mai Laboratory for Petroleum & Climate, Chiang Mai University.